\documentclass[twocolumn,showpacs,showkeys,preprintnumbers]{revtex4}%
\usepackage{amssymb}
\usepackage{makeidx}
\usepackage{amsmath}
\usepackage{graphicx}
\usepackage{dcolumn}
\usepackage{bm}
\usepackage{amsfonts}%
\providecommand{\U}[1]{\protect\rule{.1in}{.1in}}
\begin{document}
\title{Current-induced Magnetoexcitons in Mesoscopic Electron-hole Plasma}
\author{Yu. A. Pusep$^{1}$, M. A. T. Patricio$^{1}$, G. M. Jacobsen$^{2}$, M. D.
Teodoro$^{2}$, G. M. Gusev$^{3}$, and A. K. Bakarov$^{4}$}

\begin{abstract}
A radical restructuring of the optical response of highly excited
electron-hole plasma formed in a mesoscopic GaAs/AlGaAs channel in a
quantizing magnetic field when an electric current flows in the channel has
been discovered. In the absence of current, the emission spectra are caused by
transitions between Landau levels formed in the conduction band and in the
valence band of heavy holes. A critical electric current leads to a drastic
change in the emission spectra with a predominant contribution from light
holes. It is shown that the current causes local accumulation of light holes
due to Coulomb drag, which leads to strong electron-hole coupling and, as a
consequence, the formation of excitons and trions.

\end{abstract}
\date{04/07/2025}
\affiliation{$^{1}$S\~{a}o Carlos Institute of Physics, University of S\~{a}o Paulo, PO Box
369,13560-970 S\~{a}o Carlos, SP, Brazil.}
\affiliation{$^{2}$Physics Department, Federal University of S\~{a}o Carlos, 13565-905,
S\~{a}o Carlos, SP, Brazil.}
\affiliation{$^{3}$Institute of Physics, University of S\~{a}o Paulo, 135960-170 S\~{a}o
Paulo, SP, Brazil.}
\affiliation{$^{4}$Institute of Semiconductor Physics, 630090 Novosibirsk, Russia.}
\startpage{1}
\endpage{102}

\email{Corresponding author e-mail: pusep@ifsc.usp.br}
\maketitle

\section{Introduction}

The study of hydrodynamic electronic systems, in which momentum-conserving
electron-electron scattering dominates momentum-relaxing scattering, has
received much attention in the last two decades
\cite{vignale1999,geim2020,bandurin2016,moll2016,krebs2023,fritz2024,palm2024}%
. Most experiments published to date have demonstrated hydrodynamic electron
flow using electric transport measurements. However, very little is known
about the optical response in the hydrodynamic regime
\cite{pusep2023,pusep2024}. Meanwhile, the optical properties of
semiconductors provide important additional information that is not always
available in experiments on electric transport. In particular, these are data
related to the collective behavior of electrons, which is a fundamental
property of hydrodynamic systems. One of the simplest forms of collective
optical response is the excitonic features observed in the optical spectra of
the samples under study. An exciton in a semiconductor consists of an electron
bound to a hole created by optical excitation in the spectral vicinity of the
gap. The creation and recombination of excitons is one of the fundamental
mechanisms by which light interacts with semiconductors. Furthermore,
excitonic systems reveal a number of unusual properties, such as the ability
to transport energy without transporting charge. Under certain conditions,
excitons exhibit a wide range of collective properties; among them are the
formation of excitonic complexes such as biexcitons
\cite{lampert1958,moskalenko1958}, trions \cite{bar-joseph2005} and exciton
condensation \cite{butov1994}.

Another remarkable collective phenomenon is Coulomb drag, which refers to the
transport phenomenon between charge carriers where the motion of one type of
carrier influences the motion of another type due to the Coulomb interaction
between them. The electron-hole (e-h) drag was observed in Ge
\cite{mclean1961}, in Si \cite{marohashi1985} and in GaAs/AlGaAs
heterostructures \cite{hopfel1988}, Later, the e-h drag was discovered in
GaAs/AlGaAs bilayer heterosystems \cite{solomon1989,gramila1991}. In this
case, the drag in one layer is caused by the driving current in the other
layer and is due to the momentum transfer mediated by the Coulomb interaction
between the layers. Furthermore, it was found that Coulomb drag leads to the
coupling between electrons and holes and the formation of excitons
\cite{sivan1992,seamons2009}, biexcitons \cite{lee2009,maezono2013} and
excitonic condensation \cite{eisenstein2004}. However, in these cases excitons
were detected indirectly as the cause of the increased interlayer resistance.
An ample review of Coulomb drag effects can be found in \cite{narozhny2016}.
In hydrodynamic electron systems, the e-h drag is enhanced due to the dominant
carrier-to-carrier scattering, and this was observed in hydrodynamic
photoexcited e-h plasma in a mesoscopic GaAs/AlGaAs channel
\cite{pusep2022,pusep2024}.

The presented study reports Coulomb drag and the associated emergence of
excitons and excitonic complexes under the influence of electric current,
which are observed in e-h plasma formed in a mesoscopic GaAs channel as a
result of photogeneration at high excitation levels in a high magnetic field.
In fact, no effect of current on photoluminescence emission is expected. In
metallic systems, no influence of electric current on PL is expected since the
current is conducted by electrons near the Fermi surface, while PL is
dominated by electron states near the bottom of the conduction band. Thus,
conducting and recombining electrons are energetically separated and do not
influence each other. However, as will be shown below, under certain
conditions, electric current can cause radical changes in the PL spectra.

\section{Experimental}

The sample used in this work is a mesoscopic channel 5 $\mu$m wide and 100
$\mu$m long fabricated from a single GaAs quantum well (QW), with a thickness
of 14 nm, grown on a (100)-oriented GaAs substrate by a molecular beam
epitaxy. QW barriers were grown in the form of short-period GaAs/AlAs
superlattices. The sheet electron density and the mobility measured at the
temperature of 1.4 K were 9.1$\cdot$10$^{11}$ cm$^{-2}$ and 2.0$\cdot$10$^{6}$
cm$^{2}$/Vs, respectively. In this structure the viscous flow electron
transport was demonstrated in
Refs.\cite{gusev2018AIP,gusev2018PRB97,gusev2018PRB98}. The electric field
built in the barriers spatially separates the electrons and holes
photogenerated in the barriers, which leads to hole injection into the QW. As
a result, photogenerated holes move into the QW, where they form hydrodynamic
e-h plasma within the excitation laser spot. Details of e-h plasma formation
can be found in \cite{pusep2023}. Thus, the system under study consists of a
high-density inhomogeneous e-h plasma formed in a region determined by the
diffusion length of holes (about 5-10 $\mu$m) and a region where the
surrounding background electrons are located. DC voltage applied to the ends
of the channel causes an electric current to flow through the channel. When a
quantizing magnetic field is applied in addition to the electric current, the
formation of excitons and excitonic complexes is detected.%

\begin{figure}[ht]
\centering
\includegraphics[width=9cm]{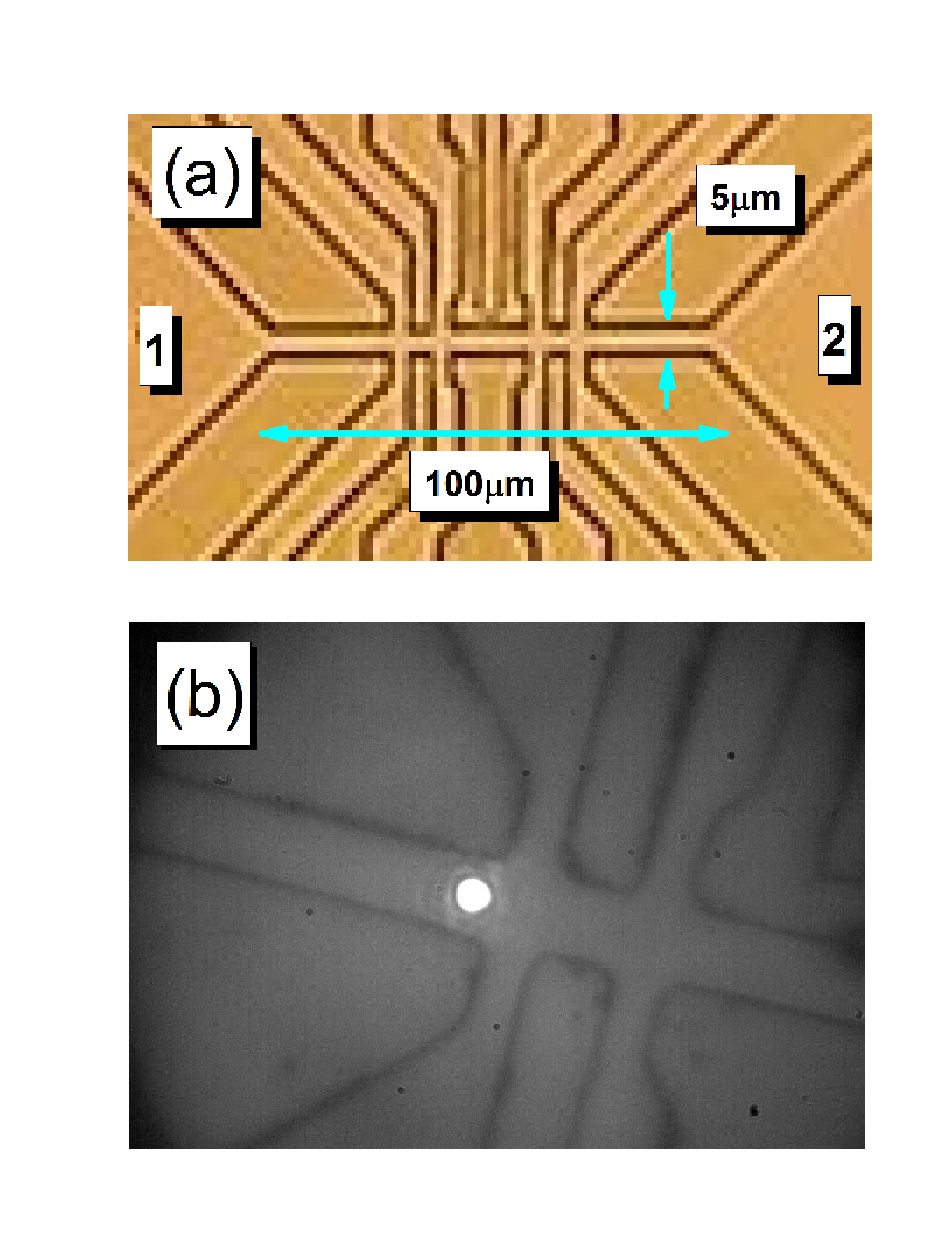}
\caption{(Color online) (a) Microscope image of the sample with characteristic dimensions. (b) Sample image with the laser spot focused on the channel.}
\label{fig:fig1}
\end{figure}

Scanning photoluminescence (PL) microscopy experiments were performed at the
temperature 4 K using a helium closed cycle cryostat equipped with a
superconducting magnet (Attocube/Attodry1000) with a magnetic field directed
perpendicular to the channel plane. PL emission was dispersed by a 75 cm
Andor/Shamrock spectrometer. The characteristic exciton recombination time was
measured by time resolved PL with a temporal resolution of 100 ps using the
Pico Quant/LDH Series diode lasers emitting 80 MHz pulses at 440 nm (2.82 eV)
with a pulse duration of 70 ps. The PL decay transients were detected by a
PicoQuant Hybrid PMT detector triggered with a time correlated single photon
PicoQuant/PicoHarp 300 counting system. E-h pairs were generated inside a
laser spot about 1 $%
%TCIMACRO{\U{3bc} }%
%BeginExpansion
\mu
%EndExpansion
$m in size, focused in the middle of the channel. The spatial resolution of
the setup is determined by the size of the light collection area, which is
estimated at about 10 $%
%TCIMACRO{\U{3bc} }%
%BeginExpansion
\mu
%EndExpansion
$m in the spectral range of GaAs QW radiation due to chromatic aberration.

\section{Results}

The PL spectra measured from the GaAs channel without a magnetic field are
shown in Fig. 2(a). PL emission is observed in the energy range between the
band gap and the Fermi level energy. The Fermi level energy extracted from the
PL spectra is found in good agreement with the Fermi level energy determined
by the magnetotransport measurements (about 30 meV). The weak double peak at
about 1.515 eV is due to recombination involving DX centers (1.5141 eV) and a
free exciton X (1.5153 eV) in the 100 nm wide buffer GaAs layer
\cite{pavesi1994}. From a comparison of the PL spectra measured at T = 4 K and
a current I$_{12}$ = 100 $\mu$A with the spectra measured at T = 30 K without
a current, it follows that such an electric current causes heating of
electrons to a temperature of about 30 K. This conclusion will be important
for explaining further results.%

\begin{figure}[ht]
\centering
\includegraphics[width=9cm]{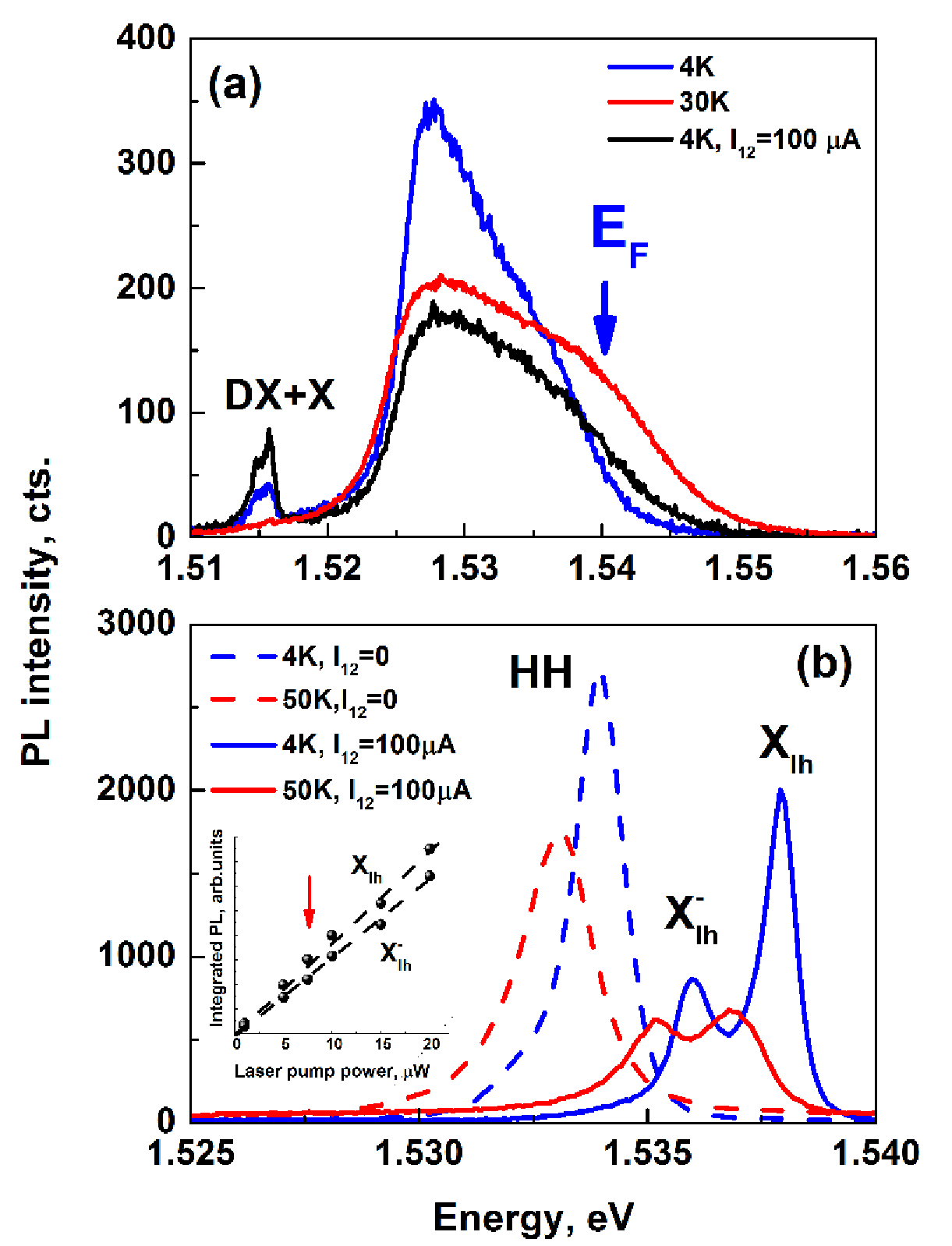}
\caption{(Color online) (a) PL spectra measured from the GaAs channel at different temperatures without current and at 4~K with a current of 100~$\mu$A. (b) PL spectra in a magnetic field of 9~T in $\sigma^{-}$ polarization at different temperatures, with and without current. Inset: Integrated PL intensity of X$_{lh}$ and X$_{lh}^{-}$ lines versus laser pump power at 4~K and 100~$\mu$A. The red arrow indicates the pump power used in the experiments.}
\label{fig:fig2}
\end{figure}

The temperature of an electron system is set by the balance between Joule
heating and thermal relaxation, determined by the energy relaxation time
$\tau_{E}$. The first law of thermodynamics states that%

\begin{equation}
\delta E=dQ+\mu dN-pdV \label{1}%
\end{equation}
where $E$ is the total internal energy, $Q$ is the heat transferred to the
system, $\mu$ is the chemical potential, $N$ is the total number of particles,
$p$ is the pressure, $V$ is the volume. With the concentration of particles
conserved $N$ = const, $V$ = const, the total energy is%

\begin{equation}
dE=dQ \label{2}%
\end{equation}

The heat transferred to electrons is $dQ=C_{e}dT=j\varepsilon\tau_{E}$, where
$C_{e}$ is the electron heat capacity, $j$ is the current density, and
$\varepsilon$ is the electric field. The change in the temperature of the
electron system $\Delta T=T_{e}-T$, where $T_{e}$ and $T=$ 4 K are the
electron temperature and the phonon bath temperature, respectively. Thus,
knowing the electron temperature, one can determine the energy relaxation time%

\begin{equation}
\tau_{E}=\frac{C_{e}\Delta T}{j\varepsilon} \label{3}%
\end{equation}

The heat capacity of two-dimensional electrons in GaAs/AlGaAs heterostructures
was measured in \cite{grivel1998} and is $C_{e}=4.14\cdot10^{-11}$ J/K$\cdot
$m$^{2}$. With the area and resistance of the channel 100$\cdot$5 $\mu$m$^{2}$
and 150 Ohm, respectively, we get $\tau_{E}=$ 0.35 ps, which is in good
agreement with the energy relaxation times $\tau_{E}=$ 1 ps and $\tau_{E}=$
0.55 ps measured in GaAs \cite{asket1968} and in GaAs/AlGaAs QWs
\cite{ozturk1992}, respectively.

The effect of the quantizing magnetic field measured in the unbiased channel
is shown in Fig. 2(b) by dashed lines. Optical transitions between Landau
levels (LL) formed in the conduction band and the valence band of heavy holes
exhibit a narrow PL line corresponding to the lowest LL. An increase in
temperature leads to a broadening of the line and its red shift due to a
decrease in the gap. A DC voltage applied to contacts 1 and 2 causes a current
I$_{12}$ to appear in the channel, which drastically changes PL: a blue
shifted double PL line emerges.

The contour plot of the PL emission measured from an unbiased channel as a
function of the magnetic field, is presented in Fig. 3(a). In GaAs the
$\sigma^{-}$ polarized PL with contribution from the spin-polarized electron
LL with the magnetic quantum number m$_{e}$ =1/2 is expected at an energy
lower than the $\sigma^{+}$ polarization, which involves electron states with
m$_{e}$ =-1/2. Indeed, the PL peak position corresponding to the lowest LL is
observed to be approximately 1 meV lower in the $\sigma^{-}$ polarization
compared to the $\sigma^{+}$ polarization. The best fit of the calculated
Landau fan to the experimental one was obtained with the exciton effective
mass 0.059m$_{0}$ which agrees well with the heavy hole exciton mass
0.06m$_{0}$ \cite{vulgarftman2001}. At the magnetic field higher than 5T only
the lowest LL is populated. In both circular polarizations, a weak PL line,
designated LH, is observed, corresponding to transitions involving light holes
\cite{ancilotto1987}.%
\begin{figure}[ht]
\centering
\includegraphics[width=9cm]{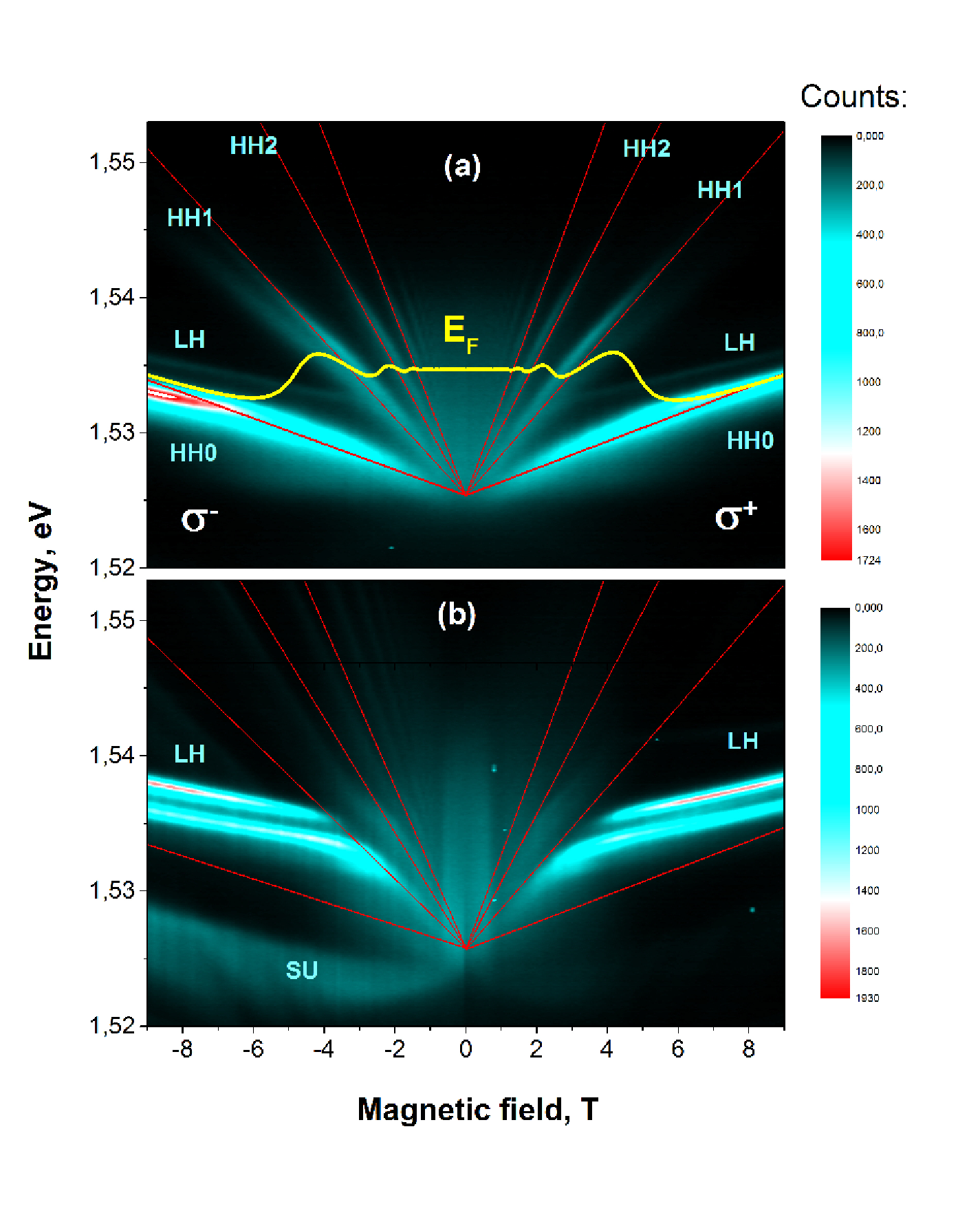}
\caption{(Color online) Contour plot of PL measured from an unbiased channel (a) and with a current of 100~$\mu$A (b) as a function of magnetic field at T = 4~K. Red and yellow lines show calculated Landau fan energies and Fermi level, respectively.}
\label{fig:fig3}
\end{figure}

The contour plot of the PL radiation from the channel measured with a current
of I$_{12}$ = 100 $\mu$A as a function of the magnetic field is shown in Fig.
3(b). Optical transitions between electron and heavy hole LLs are observed in
a magnetic field less than 2 T. In a stronger magnetic field, transitions
involving light holes dominate, which in the spectra are represented by two PL
peaks separated by an energy of about 2 meV. Moreover, at lower energy, the
weak PL emission, denoted as SU, becomes visible in $\sigma^{-}$ polarized PL,
which is due to a shake-up process involving electronic states with magnetic
quantum number m$_{e}$ = +1/2 of the lowest spin-polarized LL \cite{nash1993}.

The contour plot of the PL radiation measured as a function of current in the
channel at different magnetic fields is shown in Fig. 4. At the critical
current, indicated by the vertical arrow, PL due to transitions involving
light hole states dominates. The higher the magnetic field, the lower the
current, which leads to the predominance of light hole emission.%

\begin{figure}[ht]
\centering
\includegraphics[width=9cm]{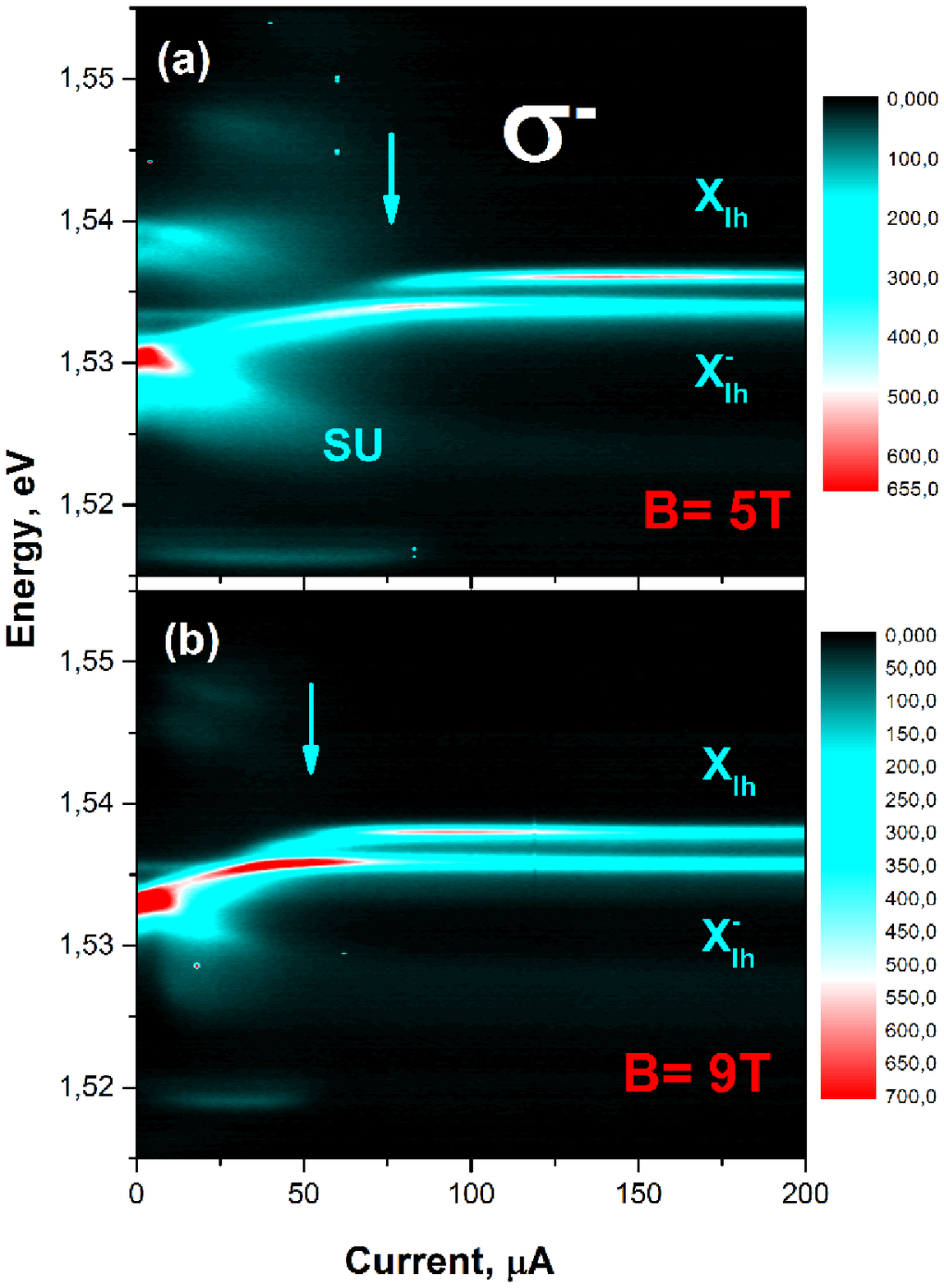}
\caption{(Color online) Contour plot of $\sigma^{-}$ polarized PL from the channel as a function of current at a magnetic field of 5~T (a) and 9~T (b) at T = 4~K. Arrows indicate the critical current above which excitons and trions are formed.}
\label{fig:fig4}
\end{figure}
The double PL line arising in a quantizing magnetic field at a critical
current, shown in Figs. 3 and 4, exhibits features of an exciton (higher
energy line) and an exciton complex (low energy line). The energy separation
between the doublet lines of 2 meV is consistent with a binding energy of 1
meV between the two excitons forming a biexciton \cite{miller1982,miller1985}
and with the binding energy of the second electron to the trion complex of
about 2 meV \cite{bar-joseph2005}. In an n-doped QW the most probable exciton
complex is a negatively charged exithon (trion). Indeed, the following
arguments support the trion version. First, in the ground state the negatively
charged trion X$^{-}$ two electrons are in a singlet state with opposite
spins. Consequently, the X$^{-}$ emitts light in both polarizations, as
observed in Fig. 3. Second, recombination of a negatively charged trion
consists of two processes: the recombination of an electron and a hole which
is accompanied by a shake-up process, when the second electron is ejecting to
one of the higher LLs \cite{finkelstein1996,finkelstein1998}. Thus the
observation of shake-up lines in Fig. 3 is important evidence for a negatively
charged trion. Moreover, the linear dependence of the integrated intensity of
the X$_{lh}^{-}$ line, shown in the inset to Fig. 2(b), indicates that a more
probable origin of the observed exciton complex is a trion rather than a
biexciton, which is characterized by a quadratic dependence.

Hence, we conclude that the double PL line in Figs. 3 and 4, which emerges
with increasing current through the channel, is caused by light hole excitons
and excitonic complexes, indicated in figures as X$_{lh}$ and X$_{lh}^{-}$, respectively.

In the reported experiments, hole injection takes place within the laser spot
area (about 1 $\mu$m), and the PL response is detected from a light collection
region of about 10 $\mu$m. The photoinjected electrons and holes diffuse from
the laser spot to the periphery of the collection region. As a consequence, PL
emission carries information about their diffusion in the channel. When an
electric field is applied to the channel, electrons flow through the diffusion
region and cause hole Coulomb drag. In this case, drifting electrons hinder
the diffusion of light holes moving in the direction opposite to the current,
while the diffusion of heavy holes remains unchanged. Electrons mainly cause
the drag of light holes, which have an effective mass close to that of
electrons and can therefore be dragged more efficiently than heavy holes. This
results in the accumulation of light holes in the region where the current
enters the PL collection area. The predominant Coulomb Hall current-induced
drag of light holes by electrons, leading to their spatial accumulation, was
observed in e-h plasma formed by electrons, light and heavy holes in
\cite{pusep2024b}. In that case, the low Hall current was insufficient to form
a high-density electron-hole plasma. Therefore, exciton complexes were not
observed. The experiments presented here were performed with an electrically
biased channel, which allowed us to implement a strong current in the channel
and, as a consequence, to observe the formation of a high-density
electron-hole plasma and accompanying exciton complexes.

A necessary condition for Coulomb drag is charge density fluctuations
\cite{zheng1993}, which in this case are provided by the heating of electrons,
discussed above. Heating of electrons increases their density fluctuations
when $k_{B}T_{e}$ $\geqslant E_{F}$. The Fermi level energy calculated
according to \cite{endo2008} as a function of the magnetic field is shown in
Fig. 3(a). The LL broadening was estimated using the spectroscopic LL
linewidth PL as $\Gamma$=1 meV. In this way, the Fermi level energies of 4 and
2 meV were determined at magnetic fields of 5 and 9 T, respectively. Assuming
that $\tau_{E}$ and $C_{e}$ are independent of magnetic field, from Eq. (3) we
obtain $I^{2}\sim E_{F}$. At 5T, excitons arise at a critical current of 75
$\mu$A. Accordingly, at 9T, the emergence of excitons is expected at a
critical current of 53 $\mu$A, which is in excellent agreement with the data
presented in Fig. 4(b). It is worth noting that the drag effect appear in the
magnetic field higher than 3 T. In accord with \cite{pusep2024a}, such a
magnetic field causes a break down of electron hydrodynamics. Thermal charge
density fluctuations are enhanced when electrons become diffuse, which
promotes the e-h drag necessary for hole accumulation.

Further evidence for the exciton nature of the double PL line was obtained by
measuring the recombination time as a function of temperature and channel
current, as shown in Fig. 5. Similar results were obtained for the
recombination times associated with the X$_{lh}$ and X$_{lh}^{-}$ lines.%

\begin{figure}[ht]
\centering
\includegraphics[width=9cm]{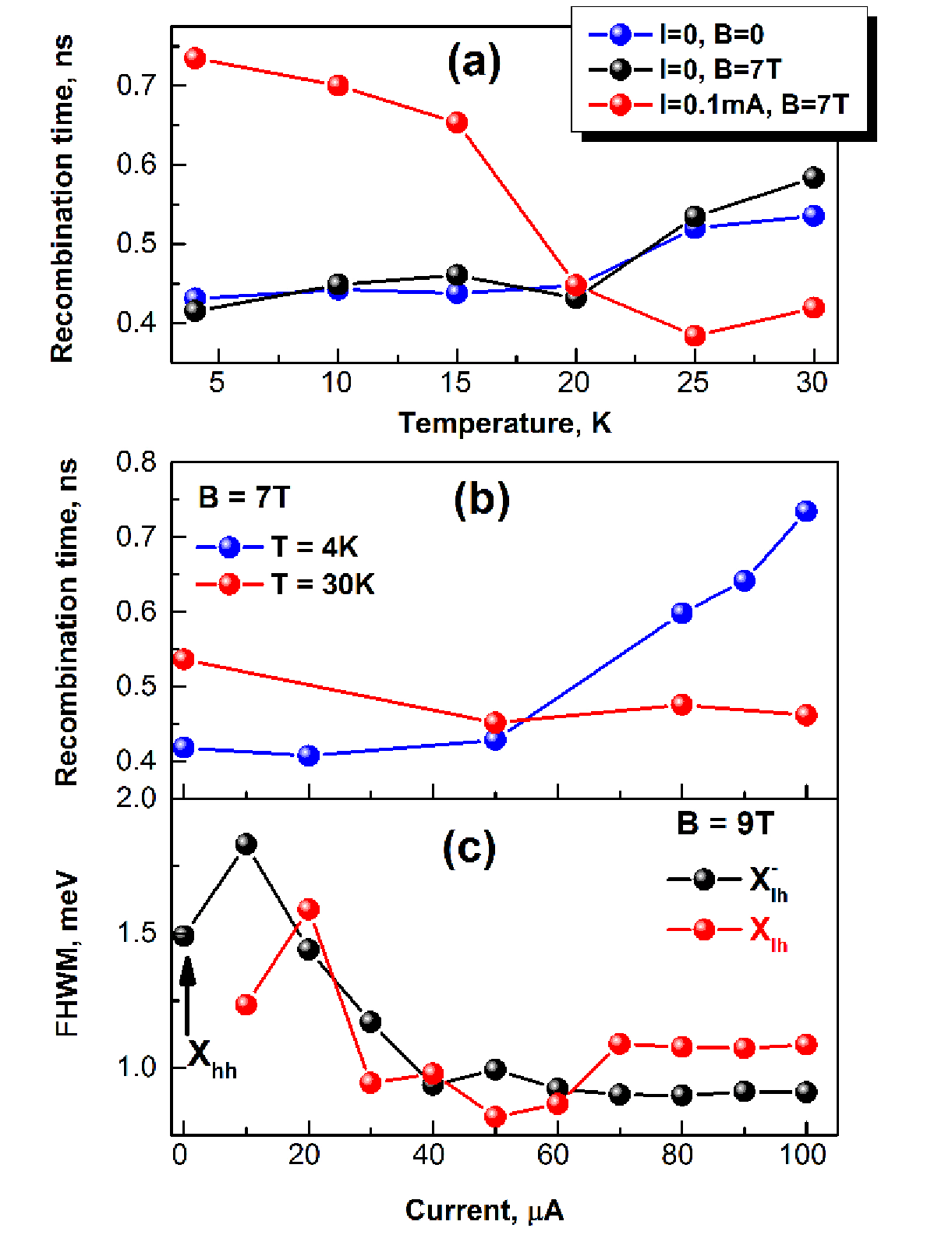}
\caption{(Color online) (a) Recombination time measured at the higher energy $\sigma^{-}$ polarized PL peak as a function of temperature. (b) Recombination time as a function of current. (c) Full width at half maximum (FWHM) of $\sigma^{-}$ polarized PL lines as a function of current.}
\label{fig:fig5}
\end{figure}
Fig. 5(a) shows that without current in the channel, both with and without a
magnetic field, the recombination time increases slightly with temperature due
to thermal excitation of carriers \cite{hooft1985}. An electric current passed
through the channel leads to a drastic change in the temperature dependence of
the recombination time. In this case, the recombination time decreases
significantly with temperature, which is consistent with the formation of
excitons at low temperature and their thermal dissociation. The effect of
current on the formation of excitons is shown in Fig. 5(b): at a temperature
of 4K, an increase in current leads to an increase in the recombination time
due to the formation of excitons, whereas at 30K, a practically constant
recombination time indicates the absence of excitons.

The width of PL lines X$_{lh}$ and X$_{lh}^{-}$, obtained in a quantizing
magnetic field using Gaussian fits, plotted in Fig. 5(c) as a function of
current, shows a broadening of both exciton states, decreasing with increasing
current. In a quantizing magnetic field without a current in the channel, the
observed PL is due to a single broad line caused by the recombination of a
heavy hole excitons X$_{hh}$. The current causes accumulation of light holes
and dominant emission of light hole excitons X$_{lh}$ and trions X$_{lh}^{-}$
with narrower linewidths.

Thus, the results presented in Fig. 5 demonstrate the transformation of
excitons in a metallic system under the action of a quantizing magnetic field
and electric current.

An important issue that needs to be discussed is the excistence of excitons
and their complexes in a heavily doped QW. At high electron density, the
screening Coulomb interaction leads to the collapse of excitons and the
exciton Mott transition \cite{mott1968}. It has been found that in GaAs/AlGaAs
QW, in a zero magnetic field the critical electron concentration above which
excitons do not exsist is about 7$\cdot$10$^{11}$ cm$^{-2}$ \cite{hisao1989},
which is close to the calculated critical Mott concentration of about
10$^{12}$ cm$^{-2}$ \cite{manzke2012}. The effect of a magnetic field is to
shrink the wave function of the electronic bound state, which leads to an
increase in the exciton binding energy and, consequently, to an increase in
the critical electron density. Moreover, the density of exciton complexes
increases and that of excitons decreases with increasing electron
concentration \cite{yoon1997}. As a result, with an increase in electron
density, the precursors of the Mott transition are exciton complexes. In the
quantum well under study, the electron concentration is close to the expected
critical Mott concentration. In this case, the magnetic field can create
conditions for the formation of exciton complexes in the QWl we are studying.

An important point is that excitons arise in a critical magnetic field at a
critical current. The critical magnetic field is probably determined by the
size of excitons, which decreases with increasing magnetic field. When the
exciton size becomes smaller than the QW width, the influence of the interface
roughness disorder decreases significantly, which favors the formation of
exciton complexes. Due to the almost identical effective mass of electrons and
light holes in GaAs, the size of the corresponding excitons in a magnetic
field B is determined by the magnetic length $L_{b}$ $=$ $\sqrt{\hbar/eB}$,
which is equal to the width of the QW of 14 nm at B = 3.4 T. This is precisely
the critical magnetic field at which light hole excitons and trions arise. At
the same time, the current performs two principal functions: it heats the
electrons and leads to a high local concentration of light holes due to
Coulomb drag, which promotes the formation of excitons and exciton complexes.
At the same time, the magnetic field quenches the kinetic energy and disrupts
the hydrodynamics of electrons, creating conditions for Coulomb drag at a
lower critical current. In addition, the magnetic field increases the exciton
binding energy and decreases the exciton radius, bringing the system closer to
the strong-coupling regime.

\section{Summary}

Finally, we conclude that the reported phenomenon of current-induced excitons
and biexcitons is caused by two factors: current-induced heating of electrons
and Coulomb drag, which entrains predominantly light holes and leads to their
local accumulation. The increased concentration of holes creates conditions
for the formation of excitons and biexcitons in the metallic electron system,
which were observed at a critical current in a highly photoexcited e-h plasma
in a quantizing magnetic field. The observed phenomenon opens up new
possibilities for the formation of high-density electron-hole plasma, where
exciton complexes can be realized.

\textbf{Acknowledgments:} YAP thanks for many valuable discussions with M. M.
Glazov. \textbf{Finding: }Financial supports from the Brazilian agencies
FAPESP (Grants 2021/12470-8, 2022/10340-2) are gratefully acknowledged.

\end{document}